\documentclass[%
 amsmath,amssymb,aps,prb,
 reprint,superscriptaddress,%
]{revtex4-1}

\usepackage{graphicx}
\usepackage{dcolumn}
\usepackage{bm}
\usepackage{hyperref}

\hypersetup{
	colorlinks = true,
	pdftitle = {},
	pdfauthor = {},
	pdfkeywords = {},
	linkcolor = blue,
	citecolor = blue,
	filecolor = black,
	urlcolor = magenta
}

\newcommand{\dd}{\mathrm{d}}

\begin{document}

\title{Quantum-mechanical treatment of atomic resolution differential phase contrast imaging of magnetic materials}%

\author{Alexander Edstr\"{o}m}
\affiliation{Materials Theory, ETH Z\"{u}rich, Wolfgang-Pauli-Stra\ss{}e 27, 8093 Z\"{u}rich, Switzerland}
\author{Axel Lubk}
\affiliation{Institute for Solid State Research, IFW Dresden, Helmholtzstra\ss{}e 20, 01069 Dresden, Germany}
\affiliation{Institute for Solid State and Materials Physics, TU Dresden, Germany}
\author{J\'{a}n Rusz}%
\email{jan.rusz@physics.uu.se}
\affiliation{Department of Physics and Astronomy, Uppsala University, P.O. Box 516, 75120 Uppsala, Sweden}%

\date{\today}

\begin{abstract}
Utilizing the Pauli equation based multislice method, introduced in Phys. Rev. Lett. \textbf{116}, 127203 (2016), we study the atomic resolution differential phase contrast (DPC) imaging on an example of a hard magnet FePt with in-plane magnetization. Simulated center of mass pattern in a scanning transmission electron microscopy (STEM) experiment carries information about both electric and magnetic fields. The momentum transfer remains curl-free, which has consequences for interpretation of the integrated DPC technique. The extracted magnetic component of the pattern is compared to the expected projected microscopic magnetic field as obtained by density functional theory calculation. Qualitative agreement is obtained for low sample thicknesses and a suitable range of collection angles.
\end{abstract}


\maketitle

\section{Introduction}

Differential phase contrast (DPC) imaging is a transmission electron microscopy (TEM) technique that measures deflections of an electron beam due to electric and magnetic fields in a thin sample \cite{dekker,chapman}. By scanning a convergent electron probe over the sample in STEM mode, spatially resolved maps of these fields are generated with this technique, which makes it an important characterization tool for nanoscale solid state phenomena. STEM-DPC has been used, for example, to detect magnetic field in magnetic domains \cite{Lee2017,Chen2018} or skyrmions \cite{Matsumoto2016,Schneider2018}, and electric fields at nano-scale and recently even at atomic resolution \cite{Shibata2012,Muller2014,Shibata2017b,Yucelen2018,Hachtel2018}.

A tilt in real space corresponds to a shift in the far field (Fourier space) and that is typically evaluated from signals detected by a four-quadrant detector in a diffraction plane. Detectors with more segments or increasingly fast pixelated detectors have been used in STEM-DPC more recently \cite{Pennycook2015,Yang2015,Krajnak2016,Shibata2017a,Brown2017,Cao2018}. With these one can acquire the whole diffraction pattern (or ronchigram) at every scan point and obtain the center of mass (COM) from each diffraction pattern numerically by post-processing \cite{Muller2014,Hachtel2018}.

Theoretical understanding of DPC is based on Ehrenfest's theorem \cite{Muller2014, Lubk2015}. The main result of these considerations is that a straight-forward  interpretation of deflection angles in terms of projected electric and magnetic fields is only possible in the phase grating approximation, i.e., for weakly scattering samples in the absence of dynamical effects. As this requirement typically gets violated under atomic resolution conditions and specimen thicknesses above a few nanometers already, a quantum-mechanical treatment of measuring electric fields by DPC at atomic resolution has been discussed thoroughly in several works \cite{Muller2014,Close2015,Muller2017,Seki2017}. 

Note, however, that atomic resolution mapping of microscopic magnetic fields has neither been reported experimentally nor described theoretically. The main reason for this blind spot is the weak perturbation of the scattered wave due to atomic magnetic fields \cite{Lubk2009} rendering an experimental detection challenging with state-of-the-art TEM instrumentation. Notwithstanding further progress in terms of stability and signal-to-noise-ratio may further increase the DPC signal resolution providing access to this very intriguing regime (e.g., for studying antiferromagnetic textures).

In this work we therefore present a quantum mechanical theory of magnetic STEM-DPC utilizing the paraxial Pauli equation multislice method as introduced in Refs.~\cite{alexprl,alexprb}. We have simulated STEM-DPC of a hard magnetic material FePt with easy axis of magnetization oriented in plane. Extracted magnetic signals at acceleration voltages ranging from 60~kV up to 1000~kV are compared to the projected microscopic magnetic field, which was obtained by density functional theory and served as an input for the calculations. Section~\ref{sec:sim} describes the simulation details. Section~\ref{sec:theo} analyses the expected magnetic contrast in the STEM-DPC images following \cite{Muller2014}. Section~\ref{sec:res} summarizes the results of our simulations. In Sec.~\ref{sec:disc} we discuss qualitatively the individual terms of the paraxial Pauli equation with focus on sources of microscopic magnetic information.

\section{\label{sec:sim}Simulation details}

We have performed multislice simulations based on paraxial Pauli equation \cite{alexprl,alexprb}, equivalently written as
\begin{eqnarray} \label{eq:Pauli}
    \lefteqn{\hat{p}_z \left( \begin{array}{c} \psi_\uparrow(\mathbf{r}) \\ \psi_\downarrow(\mathbf{r}) \end{array} \right) = \frac{1}{ \hbar k + e A_z } \Big\{-\frac{1}{2}\hat{\mathbf{p}}_\perp^2 + meV} \nonumber \\
    & - & (e\mathbf{A}_\perp) \cdot \hat{\mathbf{p}}_\perp - \hbar k e A_z - e\mathbf{B} \cdot \hat{\mathbf{S}} \Big\} \left( \begin{array}{c} \psi_\uparrow(\mathbf{r}) \\ \psi_\downarrow(\mathbf{r}) \end{array} \right)
\end{eqnarray}
where $\hat{p}_z=-i\hbar \frac{\partial}{\partial z}$ is the (canonical) momentum operator in $z$-direction, $\hat{\mathbf{p}}_\perp=(\hat{p}_x,\hat{p}_y)$, and $\psi_{\uparrow\downarrow}(\mathbf{r})$ are slowly-changing envelope wave-functions along $z$ (i.e., without the quickly oscillating $e^{ikz}$ factor) for spin up and down, respectively. Furthermore, $k$ is electron wave-vector at acceleration voltage $V_\mathrm{acc}$ and $e>0 , m , \hbar$ are the elementary charge, relativistically corrected electron mass and the reduced Planck constant, respectively. The magnetic vector potential $\mathbf{A}$ and induction $\mathbf{B}$ are obtained from electronic structure calculations (see below), whereas the electrostatic potential $V$ is generated by superposing independent atomic potentials using Kirkland's parametrization \cite{kirkland}. 

The material chosen for this study is ferromagnetic FePt. FePt crystallizes in a tetragonal L$1_0$ structure (space group P4/mmm) with $a=2.71$~\AA{} and $c=3.72$~\AA{} \cite{fept}. It has a Curie temperature safely above room temperature, close to 700~K. This material is notable for its large magnetocrystalline anisotropy energy \cite{feptmae}, which makes it important for applications, such as in magnetic recording \cite{Shibata2003,Hu2011,Weller2016}. The spin density has been calculated by density functional theory using WIEN2k code \cite{wien2k} in the generalized gradient approximation of exchange-correlation effects, see Ref.~\cite{alexprb} for details. From the spin density we have evaluated the spin current density by Gordon decomposition, followed by solving the Poisson equation to obtain the magnetic vector potential $\mathbf{A}$ in Coulomb gauge (which is used throughout this work), from which the magnetic induction $\mathbf{B}=\nabla \times \mathbf{A}$ follows. Details about the procedure can be found in Ref.~\cite{alexprb}.

In Pauli-multislice calculations presented here, we have set the orientation of the FePt crystal such that the longer $c$-axis is oriented in-plane, along the $x$-axis. The macroscopic magnetization is then oriented along the $x$-axis, along the easy axis of magnetization of the material. Due to the large magnetocrystalline anisotropy, it is expected to keep the magnetization in this direction also in the presence of a sizable magnetic field, such as that typically present in an electron microscope.

\section{\label{sec:theo}Momentum transfer to paraxial electrons in electromagnetic fields}

Ehrenfest's theorem for an electron in an electromagnetic field allows to write the dynamics of the mechanical momentum operator expectation value as
\begin{equation}
\frac{\dd^2 }{\dd t^2} \left\langle \hat{\mathbf{r}} \right\rangle = \frac{\dd }{\dd t} \left\langle \hat{\mathbf{p}} + e\mathbf{A} \right\rangle = - e \left\langle \mathbf{v} \times \mathbf{B} + \mathbf{E} \right\rangle , 
\end{equation}
which is corresponds to the classical Lorentz force law for an electron travelling at velocity $\mathbf{v}$ through the fields $\mathbf{A}$, $\mathbf{B}=\nabla \times \mathbf{A}$ and $\mathbf{E}$. In Ref.~\cite{Muller2014}, relations between momentum transfer to an electron beam and the electric field in a solid were derived without taking into account magnetic fields, i.e. $\mathbf{A} = \mathbf{B}=0$. Here, we wish to generalize this description to a situation including magnetic fields. As usual, we are interested in a situation where the velocity $\mathbf{v} \parallel \hat{e}_z$ is large, so paraxial quantum mechanics can be used. Using $\frac{\dd}{\dd t} = \frac{\dd z}{\dd t} \frac{\dd}{\dd z} = v \frac{\dd }{\dd z}$, we can write 
\begin{equation} \label{eq.Ehren_mag}
\frac{\dd }{\dd z} \left\langle \hat{\mathbf{p}} + e\mathbf{A} \right\rangle_\perp = - \frac{e}{v} \left\langle \mathbf{v} \times \mathbf{B} + \mathbf{E} \right\rangle_\perp , 
\end{equation}
where $\langle \hat{O} \rangle_\perp$ denotes the expectation value of operator $\hat{O}$ in a plane perpendicular to the propagation direction (the $xy$-plane)
\begin{equation}
\langle \hat{O} \rangle_\perp (z) = \int \dd^2 \mathbf{r}_\perp \psi^* (\mathbf{r}_\perp, z) \hat{O} \psi (\mathbf{r}_\perp, z), 
\end{equation}
with $\mathbf{r}_\perp = (x,y)$.

Here an important distinction needs to be made about the left-hand side of Eq.~\ref{eq.Ehren_mag}. In actual DPC measurement we detect the intensity of the electron beam in the diffraction plane (ronchigram, far field) and then \emph{in the post-processing} we evaluate the mechanical momentum displacement vector $\mathbf{P}$ as
\begin{equation} \label{eq:pexp}
\left\langle \mathbf{P}_\perp \right\rangle = \sum_n \mathbf{P}_{\perp,n} I_n,
\end{equation}
where the sum goes over segments or pixels of the detector. Importantly, intensities $I_n$ are measured in the \emph{field-free} far field after the electron beam has propagated through the microscope optics from the exit surface of the sample to the detector. Association of $n$ with $\mathbf{P}_{\perp,n}$ is done \emph{a posteriori} by calibrating the experimental geometry and interpreting the measured intensities. 

As a consequence equation~\ref{eq:pexp} in general differs from the quantum-mechanical expectation value of canonical momentum $\hat{\mathbf{p}}_\perp = -i\hbar \nabla_\perp$ in the object exit plane, which is gauge dependent. Gauge dependence arises from the freedom of choosing gauge for magnetic vector potential $\mathbf{A} \to \mathbf{A} + \nabla \Lambda(x,y,z,t)$. Change of gauge modifies the electron beam wave-function $\psi \to \exp\{-i\frac{e}{\hbar}\Lambda\}\psi$. Instead of gauge-dependent canonical momentum $-i\hbar\nabla$ one defines mechanical momentum $\mathbf{P}=-i\hbar\nabla + e\mathbf{A}$, as in the left-hand side of Eq.~\ref{eq.Ehren_mag}, which is gauge-invariant. Note that $\left\langle \mathbf{P}_\perp \right\rangle$ defined in Eq.~\ref{eq:pexp} approximates the expectation value of mechanical momentum transfer, expressed as a quantum-mechanical expectation value of mechanical momentum $\left\langle \psi |  \hat{\mathbf{p}}_\perp + e\mathbf{A}_\perp | \psi \right\rangle$. Precision of this approximation as a function of number of segments, their geometry, and an angular coverage of pixelated detector has been analysed in Refs.~\cite{Close2015,Muller2017,Seki2017}.

Thus, the shift in the expectation value of the electron momentum as the electron scatters through a sample of uniform thickness $d$ becomes 
\begin{equation}
\Delta \left\langle \mathbf{P}_\perp \right\rangle = -\frac{e}{v} \int_0^d \left\langle \mathbf{v} \times \mathbf{B} + \mathbf{E} \right\rangle_\perp \dd z. 
\end{equation}

For thin enough samples \cite{Muller2014,Lubk2015,Close2015,Muller2017,Seki2017} one can make the approximation $I(\mathbf{r}_\perp,z) = \psi^* (\mathbf{r}_\perp,z) \psi (\mathbf{r}_\perp,z) \approx I(\mathbf{r}_\perp , 0) \equiv I(\mathbf{r}_\perp)$. That refers to a beam centered at $\mathbf{R}=0$. Electron beam centered on $\mathbf{R}$ has an intensity distribution $I(\mathbf{r}_\perp-\mathbf{R})$. We now introduce the following notations. First, averaging over $z$ is denoted by a bar above the symbol:
\begin{equation}
\mathbf{\bar{B}}(\mathbf{r}_\perp) = \frac{1}{d} \int_0^d \mathbf{B} (\mathbf{r}_\perp ,z) \dd z,
\end{equation}
and analogically for the electric field. Such averaged variables can depend only on $\mathbf{r}_\perp=(x,y)$ coordinates. Second, the convolution with intensity of the electron beam centered at position $\mathbf{R}$ is denoted by $\otimes$ in superscript:
\begin{equation}
\mathbf{\bar{B}}^\otimes(\mathbf{R}) = \int \mathbf{\bar{B}}(\mathbf{r}_\perp) I_0 (\mathbf{r}_\perp - \mathbf{R}) \dd^2 \mathbf{r}_\perp .
\end{equation}
Such convolution is also a function of $(x,y)$ coordinates only. For clarity, we will use capital $\mathbf{R}$ to denote dependence on beam position. 


With this notation, considering that $\mathbf{v}=v\hat{e}_z$, we obtain for the shift in the expectation value of $\mathbf{P}_\perp$ 
\begin{equation} \label{eq:pR}
\Delta \left\langle \mathbf{P}_\perp \right\rangle = -\frac{ed}{v} \left[ \mathbf{\bar{E}}^\otimes + v \left(-\bar{B}^\otimes_y, \bar{B}^\otimes_{x} \right) \right],
\end{equation}
where we suppressed writing explicitly the $\mathbf{R}$-dependences. Note that the magnetic contribution increases with $v$, relative to the electric field contribution, whereby magnetic effects should increase in relative strength for high energy electron beams. 


In Appendix~\ref{sec:proof} we show that the STEM-DPC pattern (i.e., a 2D vector field) of such transversal momentum transfers remains conservative (i.e., a gradient of a scalar field) even in presence of magnetic fields. An interesting question arises, what is the scalar function $\varphi$, for which $v\hat{e}_z \times \mathbf{\bar{B}} = v \nabla \varphi$. Following \cite{alexprl}, $\mathbf{B}$ can be split into a macroscopic magnetization $\mu_0 \mathbf{M}$ and a periodic part with zero average $\mathbf{B}_\text{nc}$. Similarly, the vector potential splits into non-periodic part $\mathbf{A}_\text{np}=\frac{1}{2}\mu_0 \mathbf{M}\times \mathbf{r}$ and remaining, periodic part with zero average $\mathbf{A}_\text{p}$, where $\mathbf{B}_\text{nc}=\nabla \times \mathbf{A}_\text{p}$. For the periodic part we can then write
\begin{align}
\mathbf{\bar{B}}_\text{nc}(\mathbf{r}_\perp) = & \frac{1}{d} \int_0^d \mathbf{B}_\mathrm{nc} (\mathbf{r}_\perp ,z) \dd z \\ = & \frac{1}{d} \int_0^d \nabla \times \mathbf{A}_\mathrm{p} (\mathbf{r}_\perp ,z) \dd z
\end{align}
and we directly obtain
\begin{equation}
v\hat{e}_z \times \mathbf{\bar{B}}_\text{nc} = v \nabla \bar{A}_{\text{p},z},
\end{equation}
where we have used periodicity of $\mathbf{A}_\text{p}$, whenever derivatives with respect to $z$ have appeared. For the macroscopic part of the magnetic field, which is parallel to $x$-axis in our case, the scalar potential is $v \mu_0 M y$, so that 
\begin{equation} \label{eq:Bpot}
 v\hat{e}_z \times \mathbf{\bar{B}} = v \nabla \left( \bar{A}_{\text{p},z} + \mu_0 M y \right),
\end{equation}
The apparent gauge dependence does not actually play a role here. For the reason that we work in Coulomb gauge, the magnetic vector potential is determined up to a constant vector, the gradient of which vanishes.

Equation \ref{eq:Bpot} determines the scalar potential of $v\hat{e}_z \times \mathbf{\bar{B}}$. This has consequences for interpretation of the integrated DPC (iDPC) technique \cite{Lazic2016,Yucelen2018,Hachtel2018}, because that means that in magnetic materials the extracted scalar potential will not reflect purely the electrostatic potential---it will also contain a magnetic contribution.

\section{\label{sec:res}Results}

\subsection{Projected magnetic fields}\label{sec:bfield}

\begin{figure}
	\centering
	\includegraphics[width=0.48\textwidth]{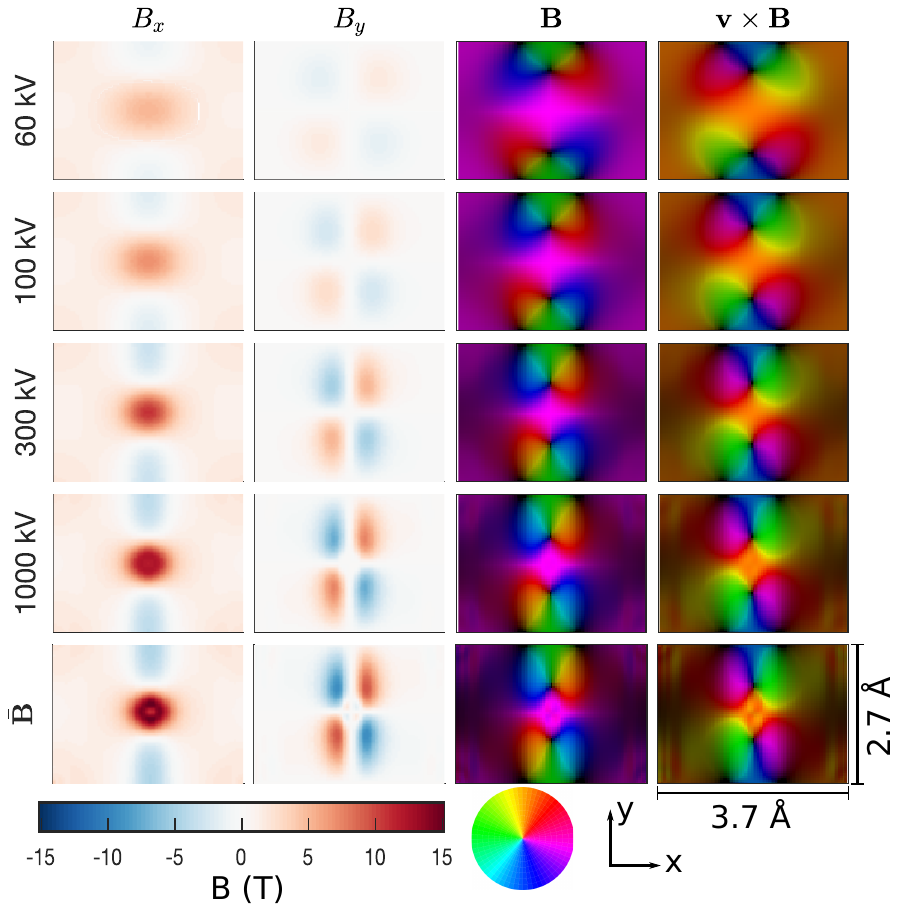}
	\caption{The magnetic field averaged in the $z$-direction $\mathbf{\bar{B}}$ (bottom row) and its convolution with the initial beam intensity, $\mathbf{\bar{B}}^\otimes$, for a beam with 25~mrad convergence angle and various acceleration voltages of 60~kV, 100~kV, 300~kV and 1000~kV  in the first to fourth rows. The fields are shown within one unit cell with the Fe atom in the middle and Pt at the corners. The first two columns show the $x$ and $y$-components of the fields. The third column shows the vector field represented with hue indicating the direction of the field and the value its magnitude. The final column shows $\mathbf{v}\times\mathbf{B}$ represented in the same way, for $\mathbf{v} = v \hat{e}_z$. The color wheel indicates the directions corresponding to the hues. }
	\label{fig.Bavg}
\end{figure}

Based on Ehrenfest theorem and derivations in Sec.~\ref{sec:theo} the magnetic contribution that we would expect to see in the STEM-DPC experiment is approximately given by the $z$-averaged magnetic induction, convolved with squared modulus of the electron beam wave-function, see Eq.~\ref{eq:pR}. Namely, the goal is to observe
\begin{equation}
    - e d \mu_0 M \hat{e}_y - ed (-\bar{B}_{\text{nc},y}^\otimes , \bar{B}_{\text{nc},x}^\otimes)
\end{equation}
The macroscopic field term is a constant throughout the unit cell and independent from the shape of the (normalized) electron beam wave-function. On the other hand, the second term originates from microscopic magnetic fields that vary within the unit cell, and therefore $\bar{B}_{\text{nc},y}^\otimes,\bar{B}_{\text{nc},y}^\otimes$ will depend on the beam shape. In this work we have kept convergence semi-angle fixed at 25~mrad, but the acceleration voltage $V_\text{acc}$ was varied between 60~kV to 1000~kV, thus the $\mathbf{R}$-dependence of the magnetic contribution will be dependent on $V_\text{acc}$. In Fig.~\ref{fig.Bavg} we have plotted the $x,y$-components of $\mathbf{\bar{B}}^\otimes$ including $\mathbf{r}_\perp$-dependence of $\mathbf{\bar{B}}$, which can be considered as a limit of $\mathbf{\bar{B}}^\otimes$ for an infinitely thin probe. Note the gradual blurring of the magnetic signal as the voltage is gradually lowered. These results will serve as a reference, to which we can compare the magnetic contribution from simulated STEM-DPC images.

\subsection{Simulated STEM-DPC images}

\begin{figure}
    \centering
    \includegraphics[width=8.6cm]{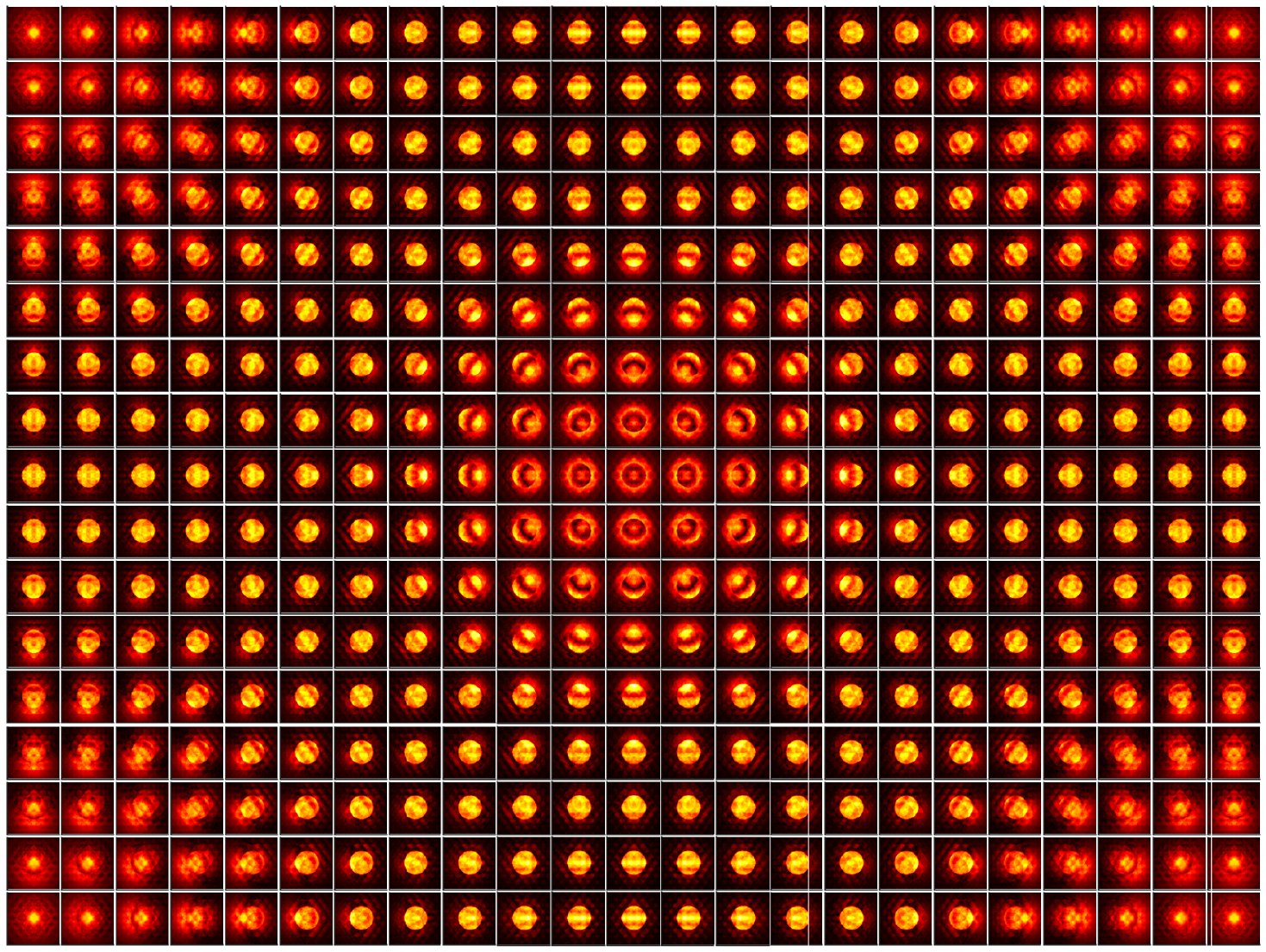}
    \caption{STEM diffraction patterns on a $23 \times 17$ grid over a Fe-centered unit cell within a 10 unit cell (2.7~nm) sample of FePt, for a beam with 100~kV acceleration voltage and 25~mrad convergence angle. Diffraction patterns are shown for a maximum collection semi-angle of 70~mrad.}
    \label{fig:difpat}
\end{figure}

Overall STEM-DPC images are dominated by the interaction of electron beam with local electric fields and those have been already analysed in detail in several works before, see Refs.~\cite{Muller2014,Close2015,Muller2017,Seki2017}. Therefore we will keep this section concise and focus more on the magnetic component of the STEM-DPC image in the following sections. In Fig.~\ref{fig:difpat} we show a composite image of the ronchigrams at all calculated beam positions for FePt sample 1.6~nm thick at acceleration voltage 100~kV and convergence semi-angle of 25~mrad. As is expected, instead of simple shifts of the central CBED disk as we scan across the unit cell, we observe the redistribution of intensity in the ronchigrams,   leading to nonzero first moments from which one can estimate the momentum transfer $\langle \mathbf{P}_\perp \rangle(\mathbf{R})$.

\begin{figure}
    \centering
    \includegraphics[width=8.6cm]{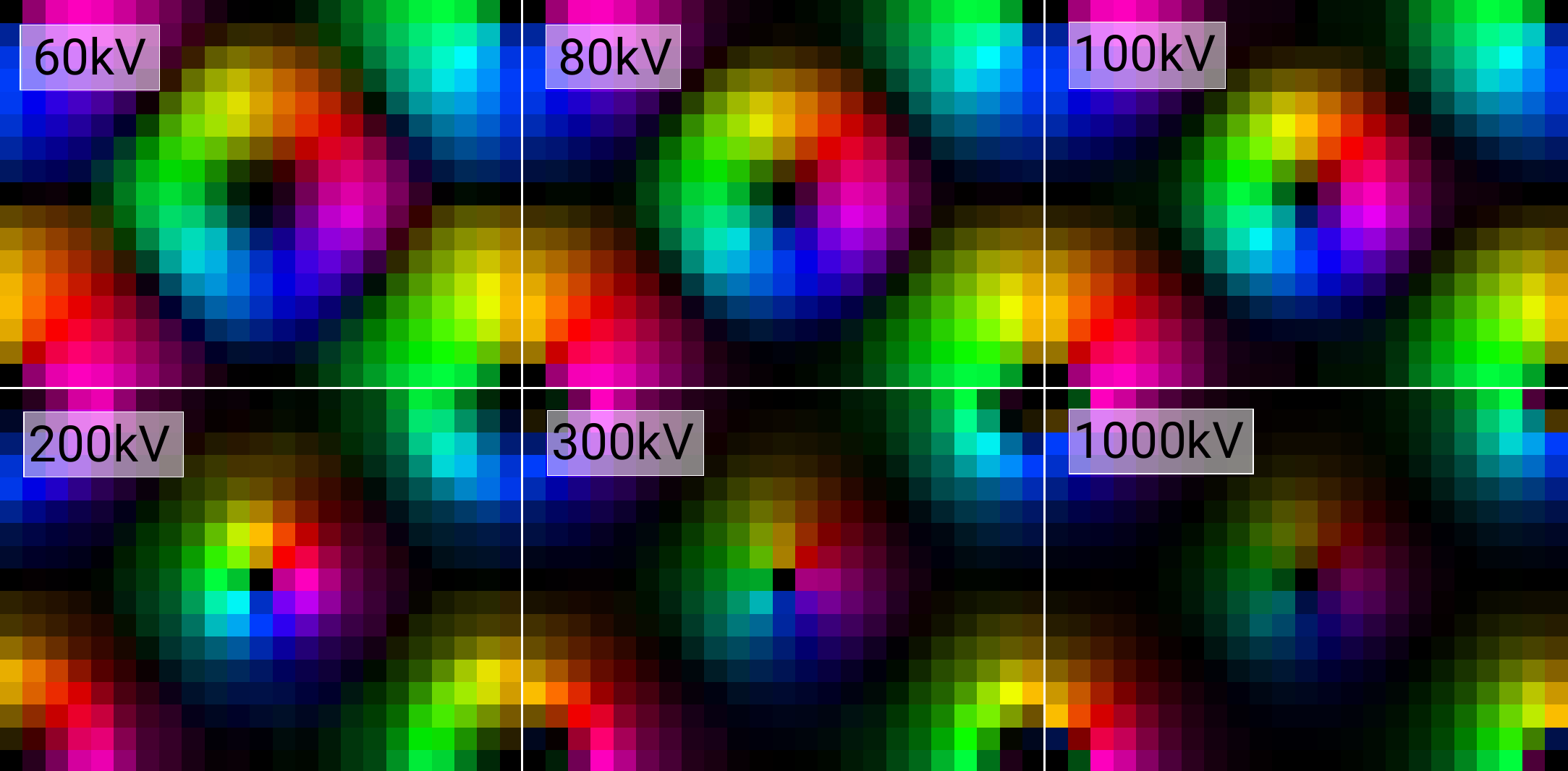}
    \caption{DPC-STEM images of FePt at thickness 2.7~nm (10 unit cells) and collection semi-angle of 45~mrad, shown for various acceleration voltages. HSV color scheme was used with hue representing direction of vectors, saturation was set to 1 and value is proportional to the length of the vector $-\langle \mathbf{P}_\perp \rangle(\mathbf{R})$, scaled to optimally use the color range.}
    \label{fig:elDPCvolt}
\end{figure}

According to Eqn.~\ref{eq:pR}, assuming that magnetic contributions are much weaker than electric ones, we should obtain an image corresponding to $\mathbf{\bar{E}}^\otimes$. Constructing such images for collection semi-angles of 45~mrad one obtains patterns shown in Fig~\ref{fig:elDPCvolt}, displayed at various acceleration voltages. In agreement with previous theoretical and experimental work at atomic resolution, we see a vector field ``emanating'' from atom positions. It is wider and more intense for the heavier Pt atom. As a function of acceleration voltage, an expected trend can be seen: with increasing voltage the beam diameter decreases and thus the DPC pattern due to Coulomb fields becomes sharper.

\subsection{Macroscopic magnetization from STEM-DPC images}

As discussed in Ref.~\cite{Muller2017}, in order to detect macroscopic electric fields that are typically several orders of magnitude smaller than the local electric fields, one needs a very accurate summation of $\Delta \langle \mathbf{P}_\perp \rangle(\mathbf{R})$ over the whole unit cell, in order to achieve cancellation of the local fields. The same is applicable here, if our goal is to extract the macroscopic magnetization. Nevertheless, within our proof-of-concept theoretical investigation, we can achieve perfect cancellation of the electric field components thanks to the symmetry of the system and a suitably chosen grid of scan points, which reflects this symmetry.

Projected electric field vectors within the $a-c$ plane in FePt have several symmetries: horizontal and vertical mirrors and a rotation by 180 degrees around the center of projected unit cell. Note that here we speak about symmetries that also correspondingly transform the directions of vectors. If we would consider separately the $x,y$-components of the electric field, then the $x$-component is symmetric with respect to the horizontal mirror and antisymmetric with respect to the vertical mirror. And for the $y$-component the situation is reversed. Both components change sign under a 180 degree rotation.

These symmetries are necessarily also reflected in the resulting COM vector field. Conveniently, under an assumption that the electric field component of the COM is proportional to the $\mathbf{\bar{E}}$, this means that summing the COMs over the whole unit cell should lead to an exact cancellation of the electric field contribution to the COMs. In context of Ref.~\cite{Muller2017}, a sufficient condition here is that the unit cell of FePt crystal has an inversion symmetry, and thus no electric polarization.

\begin{figure}
    \centering
    \includegraphics[width=8.6cm]{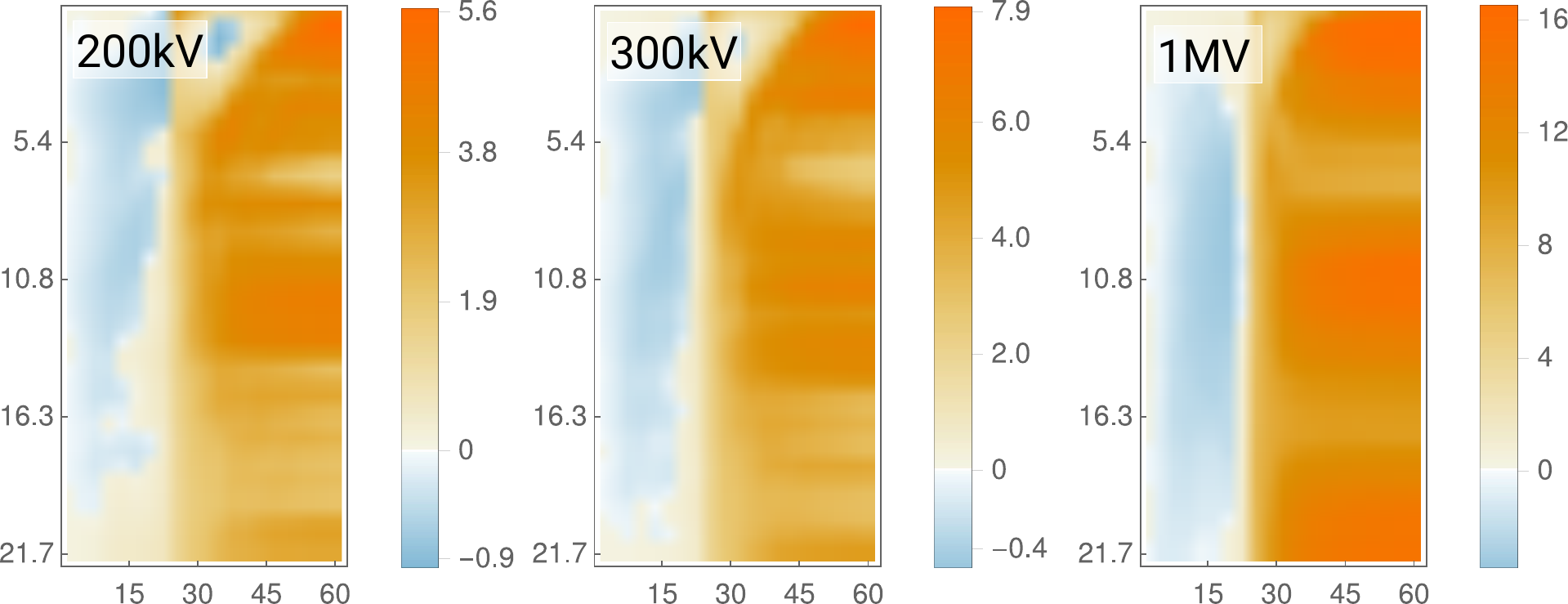}
    \caption{Unit cell average of $\frac{1}{d}\langle \mathbf{P}_\perp \rangle(\mathbf{R})$ (arbitrary units) as a function of collection semi-angle in mrad (horizontal axis), thickness in nm (vertical axis) and voltage (individual panels).}
    \label{fig:COMsum}
\end{figure}

The symmetry of projected magnetic field components is lower. While the $x$-component of magnetic field is symmetric with respect to horizontal and vertical mirrors, the $y$-component is antisymmetric with respect to both. On the overall only the 180 degree rotation remains. This difference will be utilized in the next subsection for isolating the microscopic magnetic component of COM field. Here it is sufficient to realize that the non-constant part of the magnetic field $\mathbf{B}_\mathrm{nc}$ averages to zero by construction and this property transfers to $\mathbf{\bar{B}}$ and also $\mathbf{\bar{B}}^\otimes$. Therefore when we sum COMs over the whole unit cell, eventually only the component due to constant macroscopic magnetization component should remain.

We have checked that such sum is numerically a zero, when a standard multislice calculation without magnetic fields is performed. Figure~\ref{fig:COMsum} shows, how such sum evolves as a function of acceleration voltage, sample thickness and collection semi-angle. The picture is qualitatively similar to previous analyses of the electric signal. The collection semi-angle needs to be larger than the convergence semi-angle. The averaged $\langle \mathbf{P}_\perp \rangle$ remains stable above certain collection semi-angle. For ultra-low thicknesses the collection semi-angle needs to be appreciably larger than the convergence semi-angle, though this requirement softens with increasing voltage. In addition, even at large collection angles, as a function of sample thickness we observe fluctuations of the $\frac{1}{d} \langle \mathbf{P}_\perp \rangle$ reminiscent of those reported by M\"{u}ller et al.\ for electric fields \cite{Muller2014}.

From a practical perspective, it is important to have a qualitative picture about the signal strengths that one can expect in experiments. From our simulations, the average relative strength of the DPC signal component due to the average magnetization is of the order of 0.1\% at thicknesses below 10 unit cells. This signal strength increases to approximately 0.5\% if we consider larger sample thicknesses (up to 80 unit cells, i.e., 21.7~nm). Average relative strength was here estimated as $|\sum_{i,j} \mathrm{DPC}_y(i,j)|/\sum_{i,j} ||\mathrm{DPC}_(i,j)||$, where $(i,j)$ label the grid points within a unit cell.

\subsection{Microscopic magnetization from STEM-DPC images}

As indicated above, one could use the different symmetries of the electric and magnetic component of the STEM-DPC image in order to isolate them. This is of course system dependent and not always possible, e.g., when dealing with materials of low symmetry. Nevertheless, it is applicable for FePt. 

An alternative and more general approach is to take a difference of two calculations, which differ only by changing the sign of the magnetization, here $\mathbf{M}=(M,0,0) \to (-M,0,0)$. (This approach could also be implemented in actual measurements.) In general, two separate calculations would be needed. For FePt thanks to its symmetry this can be achieved simply by rotating the unit cell by 180 degrees together with all momentum transfer vectors. Then, taking a difference of such two STEM-DPC images one should obtain the magnetic component of STEM-DPC image.

\begin{figure}
    \centering
    \includegraphics[width=8.6cm]{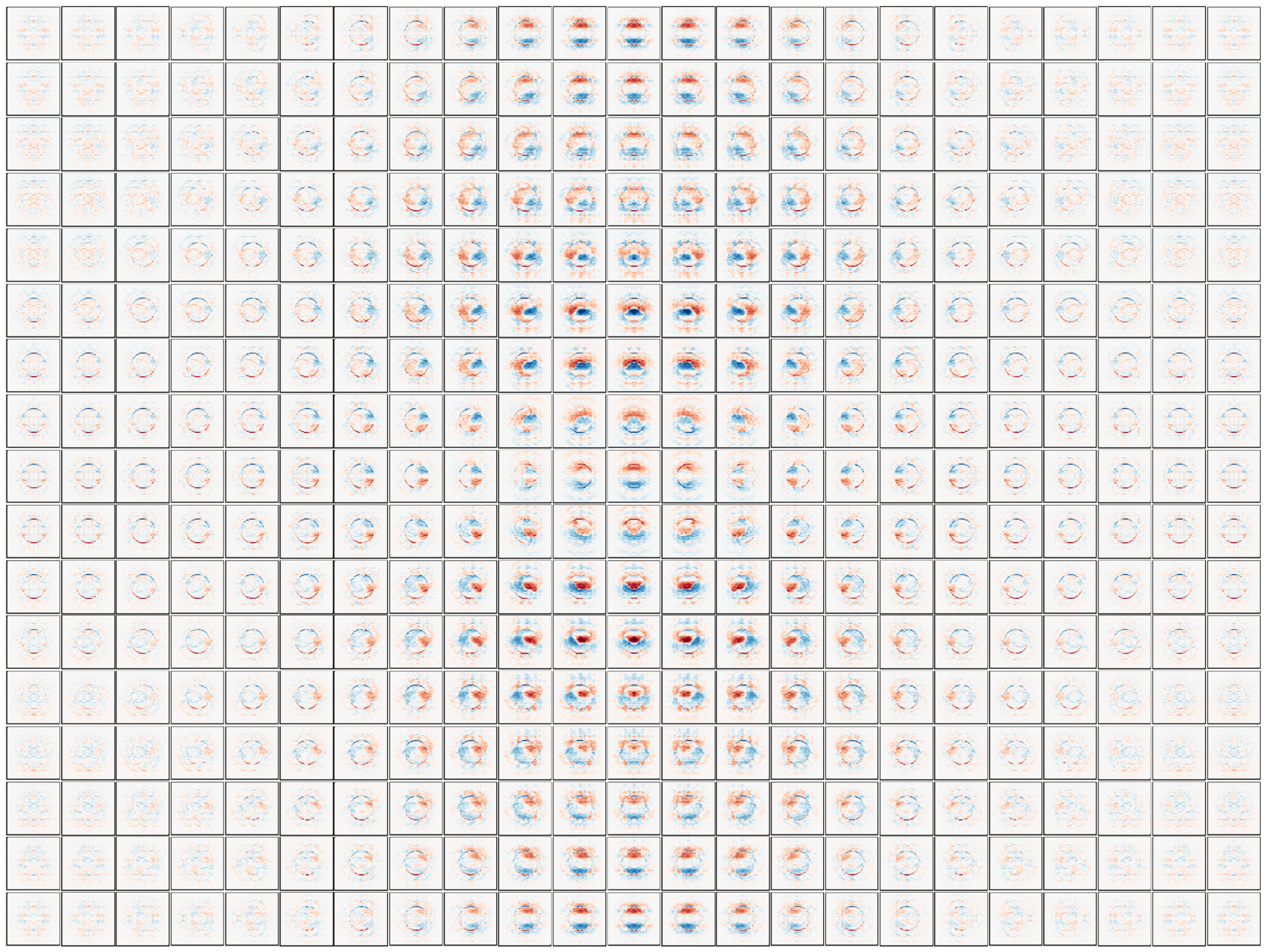}
    \caption{Magnetic component of the STEM ronchigrams in Fig.~\ref{fig:difpat}, as obtained from the symmetry arguments described in the text. Blue and red regions correspond to opposite sign contributions and the magnitudes are approximately $10^{-3}$ of those in Fig.~\ref{fig:difpat}.}
    \label{fig:V100_magDP}
\end{figure}

\begin{figure}
    \centering
    \includegraphics[width=8.6cm]{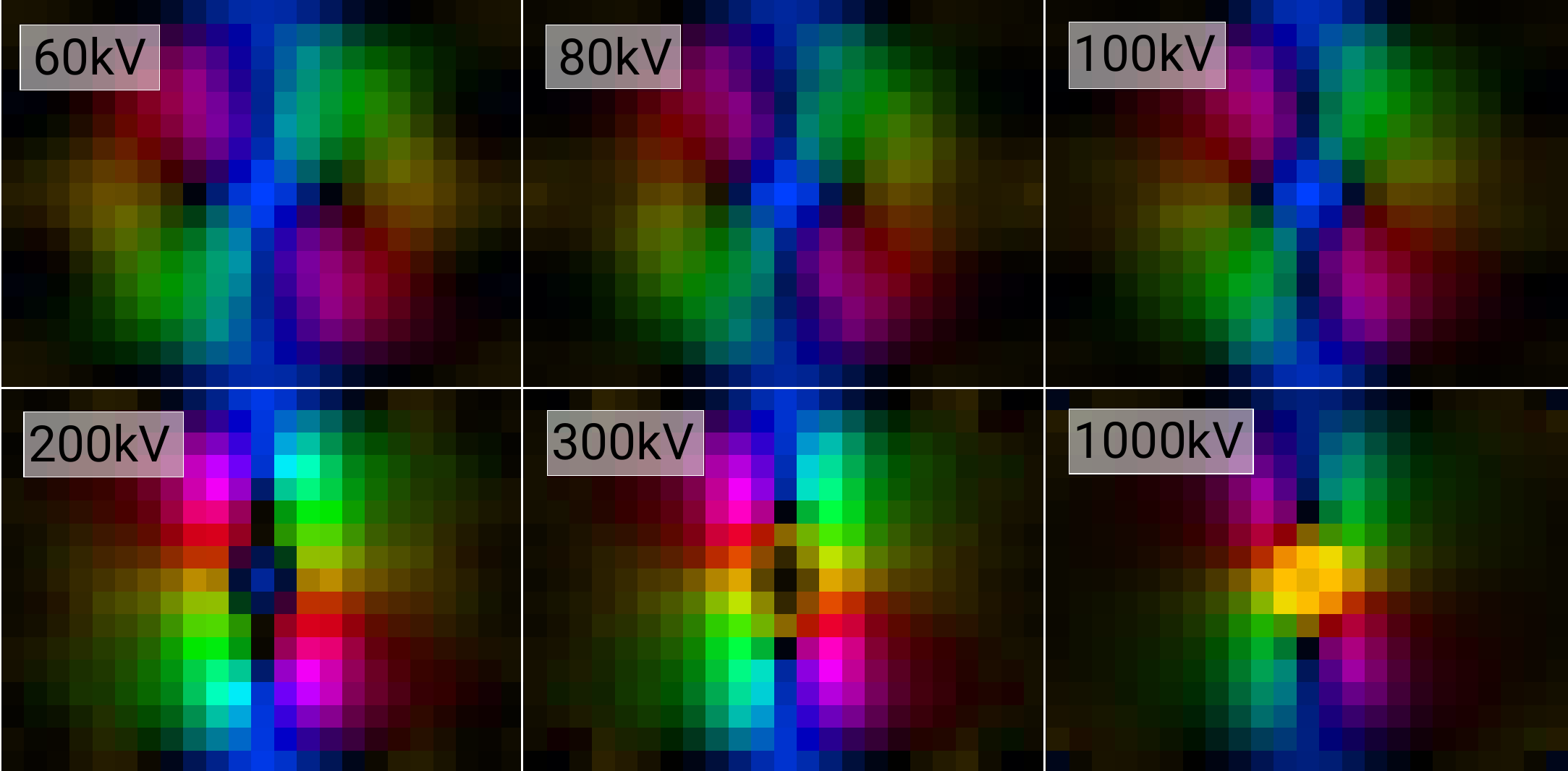}
    \caption{Magnetic component of STEM-DPC images of FePt at thickness 2.7~nm (10 unit cells) and collection semi-angle of 30~mrad, shown for various acceleration voltages. HSV color scheme was used with hue representing direction of vectors, saturation was set to 1 and value is proportional to the length of the vector $-\langle \mathbf{P}_\perp \rangle(\mathbf{R})$, scaled to optimally use the color range.}
    \label{fig:magDPCvolt}
\end{figure}

We have applied this method to extract the magnetic component of the calculated STEM-DPC images. The result of this procedure is shown in Fig.~\ref{fig:V100_magDP}, containing the differences between STEM ronchigrams in Fig.~\ref{fig:difpat}, for inverted beam positions ($\mathbf{R} \rightarrow -\mathbf{R}$) and ronchigrams ($\mathbf{k} \rightarrow -\mathbf{k}$). Inspecting this figure, one can observe the expected concentration of the magnetic signal in the neighborhood of the iron atomic column.

By calculating the COM of the data in Fig.~\ref{fig:V100_magDP}, one obtains vector fields such as those in Fig.~\ref{fig:magDPCvolt}. The images are in qualitative agreement with the $v\hat{e}_z \times \mathbf{\bar{B}}^\otimes$ shown in Fig.~\ref{fig.Bavg}. One can observe how the pattern blurs with decreasing acceleration voltage due to increasing diameter of the electron beam. Note also the yellow ``background fog'' due to the constant macroscopic magnetization component in $y$-direction.

At a closer inspection, one can spot that the $v\hat{e}_z \times \mathbf{\bar{B}}^\otimes$ is not well represented by the magnetic component of COM, especially at lower voltages. In the region where Fe atomic column is located, vectors of the reconstructed image actually point in opposite direction than the microscopic magnetic field would dictate. This can be assigned to the fragility of phase grating approximation in the atomic resolution regime, especially when looking on a weak component of the total STEM-DPC pattern. This view is supported by observing that the distortion decreases with increasing acceleration voltage. As the acceleration voltage is increasing, the scattering cross-section decreases, weakening thus the dynamical diffraction effects distorting the electron beam wave-function. Furthermore, at a sample thickness of only 6 unit cells the magnetic DPC patterns (not shown) qualitatively agree with the projected magnetic fields, even at lower acceleration voltages.

We have checked that this method of extraction of magnetic signal leads to numerical zeros, when applied to a non-magnetic multislice calculation. If we assume the linear regime, then the availability of non-magnetic calculation offers an alternative way of extraction of magnetic signal component---by a subtraction of the non-magnetic STEM-DPC pattern from the magnetic one. We have performed this test at 100kV and the result (not shown) is visually indistinguishable from what is shown in Fig.~\ref{fig:magDPCvolt}.

\begin{figure}
    \centering
    \includegraphics[width=8.6cm]{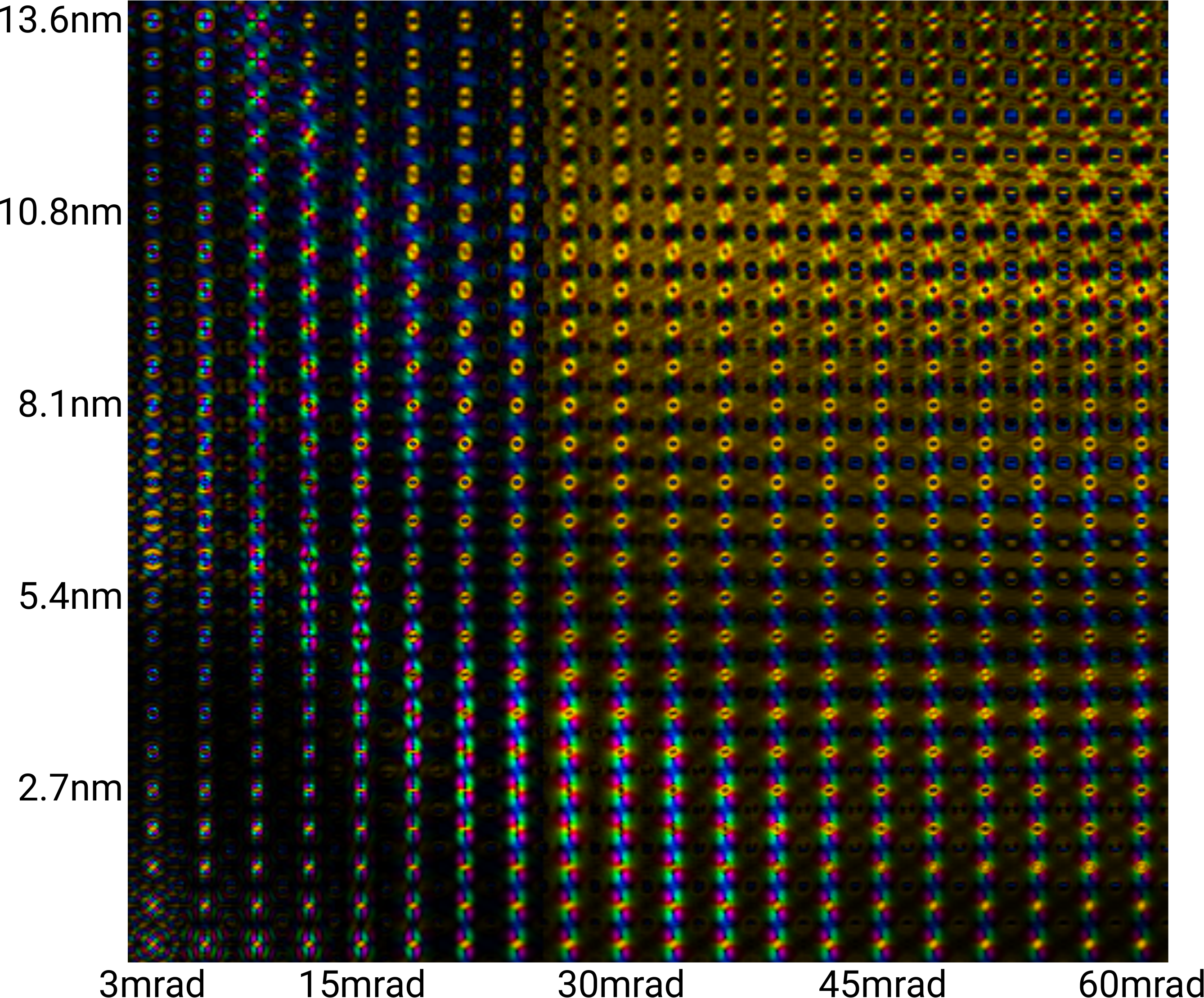}
    \caption{Magnetic component of the STEM-DPC signal as a function of sample thickness (arranged vertically) and collection semi-angle (arranged horizontally) for an acceleration voltage 300~kV.}
    \label{fig:V300tab}
\end{figure}

Figure~\ref{fig:V300tab} shows how the extracted magnetic STEM-DPC pattern depends on the collection angle and sample thickness. A strong sensitivity to both parameters is observed. Especially at higher sample thicknesses (above 8~nm) there is a clear change of the pattern, once the collection semi-angle becomes larger than the convergence semi-angle. In that region the STEM-DPC pattern is dominated by yellow color representing magnetization in $y$-direction. Nevertheless, representation of the microscopic fields in that region is not satisfactory. Those are best reproduced at low sample thicknesses below 4~nm and collection angles not far from 30~mrad. At larger collection angles, with exception of the very lowest thicknesses, the STEM-DPC pattern develops a ring-like feature at the Fe atomic column with a minimum in the center, which does not correspond to the distribution of $v\hat{e}_z \times \mathbf{\bar{B}}^\otimes$, although other features are reproduced qualitatively well.

Finally we comment on the average relative strength of the magnetic signal. This time we use the following metric $\sum_{i,j} ||\mathrm{MAGDPC}(i,j)||/\sum_{i,j} ||\mathrm{DPC}(i,j)||$, which collapses to the definition used above when the magnetic component of DPC pattern would be constant and point in $y$-direction. Intuitively one would expect that this can lead to significantly larger percentages than the average relative strength of the macroscopic magnetization components, because the local magnetic fields are substantially larger in magnitude. However, there are two effects that counter-act this intuition. First, there is a rather strong cancellation of the local magnetic fields when evaluating the $z$-averaged $\mathbf{\bar{B}}$. Second, the local magnetic fields are appreciably strong only in a relatively small part of the unit cell, nearby the iron atomic column. Both combined lead to an observation that the average relative strength of the magnetic component, as defined above, remains below 1\%, although, if we restrict the summation to a closer neighborhood of iron magnetic column, strengths of above 1\% can be observed at higher thicknesses.

\section{Discussion}\label{sec:disc}

Inspecting the paraxial Pauli equation, Eq.~\ref{eq:Pauli}, allows to analyse qualitatively how magnetism influences the electron beam wavefunction. We will focus here on the microscopic magnetization $\mathbf{\bar{B}}_\text{nc}$.

Magnetic induction $\mathbf{B}$ appears only in the last term, multiplied by a vector composed of Pauli spin matrices. For an unpolarized electron beam, as is common in transmission electron microscopes, this term doesn't allow for an extraction of local magnetic fields. Changing the sign of the magnetic moment has the same effect as inverting the spin moment of electrons. When operating a microscope with spin-polarized electron beams, this term would open for an interesting opportunity to map local magnetic fields by changing the spin polarization. We haven't pursued this option in this manuscript, being focused on standard STEM-DPC imaging, nevertheless this will be addressed in a future work.

\begin{figure}
    \centering
    \includegraphics[width=8.6cm]{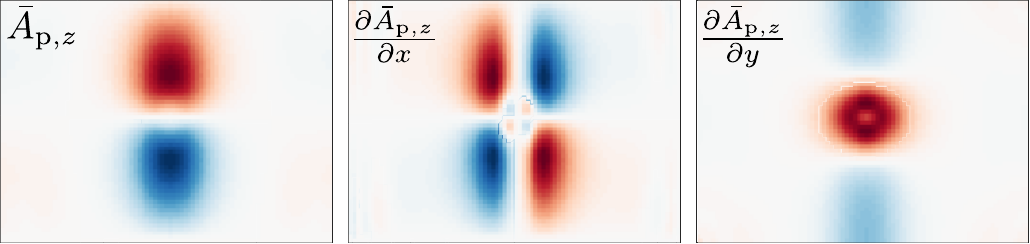}
    \caption{$z$-component of the projected magnetic vector potential $\mathbf{\bar{A}}_\text{p}$ (left) and its derivatives with respect to $x$ (center) and $y$ (right; all in arbitrary units). Platinum atomic columns are in the corners and iron atomic column is in the center of each panel.}
    \label{fig:Afield}
\end{figure}

On the other hand, magnetic vector potential $\mathbf{A}$ appears in several terms in Eq.~\ref{eq:Pauli}. First, $A_z$ appears in the denominator. Quantitatively, however, relative magnitude of $eA_z$ compared to $\hbar k$ is less than $10^{-5}$ in our simulations. Thus it is only a very weak effect. Next, the $x,y$-components of vector potential $\mathbf{A}$ appear in a term with the gradient of the electron beam wave-function. For our system, evaluation of the $\mathbf{\bar{A}}_\text{p}$ shows that its $x,y$-components are zero. Therefore in the linear regime, when the momentum transfer is proportional to the projected fields, this term doesn't contribute to microscopic magnetic signal either. This leaves the last option, the term $\hbar k e A_z/m$.

Indeed, $z$-component of $\mathbf{\bar{A}}_\text{p}$ remains nonzero and modifies the electron beam wave-function. However, its spatial distribution (see Fig.~\ref{fig:Afield}) reminds neither the distribution of the in-plane components of magnetic induction $\mathbf{\bar{B}}_\text{nc}$ nor $v\hat{e}_z \times \mathbf{\bar{B}}_\text{nc}$, see Fig.~\ref{fig.Bavg}. Nevertheless it is the term $\hbar k e A_z/m$ alone, which allows detection of projected microscopic magnetic fields. We will back-track how this happens. The average mechanical momentum transfer in $x$-direction is evaluated by $\langle \psi | \hat{p}_x - eA_x | \psi \rangle$. In the linear regime, we expect that the electron beam wave-function change due to the presence of nonzero $A_z$ term will be $|\psi_0\rangle \to |\psi_0\rangle - i \frac{ed}{\hbar}\bar{A}_z |\psi_0\rangle$. Associated change of the momentum transfer is
\begin{eqnarray}
  \delta \langle \psi | \hat{p}_x+eA_x | \psi \rangle & = & -i\frac{ed}{\hbar} \langle \psi_0 | \hat{p}_x \bar{A}_z - \bar{A}_z \hat{p}_x |\psi_0\rangle \nonumber \\
  & = & -ed \langle \psi_0 | \frac{\partial \bar{A}_z}{\partial x} | \psi_0 \rangle \nonumber \\
  & = & -ed \int \frac{\partial \bar{A}_z}{\partial x} I(\mathbf{r_\perp - R}) \dd \mathbf{r}_\perp
\end{eqnarray}
where we neglected the quadratic terms in $A_z$. Analogic relation can be derived for the $y$-component. That shows that the $A_z$ term influences the momentum transfers via its spatial derivatives. Now if one compares the $x,y$-derivatives of $\bar{A}_{\text{p},z}$ shown in Fig.~\ref{fig:Afield} to minus $y$- and plus $x$-components of $\bar{\mathbf{B}}_\text{nc}$, there is a close correspondence. Note that this is well reflected by Eq.~\ref{eq:Bpot} derived above. In addition, this analysis demonstrates from another angle of view that as $V$ is the scalar potential potential for electric fields, $v\bar{A}_{\text{p},z}$ plays the role of scalar potential of $v\hat{e}_z \times \mathbf{\bar{B}}_\text{nc}$.

Let's briefly discuss the experimental challenges involved in detecting the magnetic signal. The average magnetic signal component has typically a strength well below 1\% of the average of the electric signal component of the DPC pattern. Its isolation requires to take a difference of two separately measured pixels of data---either from a different region of the unit cell or from the same region, but after the magnetization has been inverted. Thus sample drift could cause substantial challenges. Fortunately, DPC as an integral technique uses majority of scattered electrons in evaluation of the momentum transfer. Thus it is very efficient per unit of beam current. Yet, we expect that mapping of magnetic fields will require longer dwell times, which could make the measurement more susceptible to sample drifts. Future experiments should attempt to optimize the signal to noise ratios so that the momentum transfers can be measured with a precision substantially better than 1\%, preferably though in the 0.1\% range, in order to allow extraction of the magnetic component. This might be achievable via multiframe recording \cite{Jones2018} and/or averaging the signals over a larger number of unit cells with a careful control over the scan noise.

\section{Conclusions}\label{sec:concl}

We have presented a quantum mechanical theory of differential phase contrast imaging at atomic resolution for magnetic materials. We found that even in the presence of in-plane magnetic fields, the expected momentum transfers remain curl-free. This has consequences on integrated differential phase contrast imaging in that the extracted scalar potential is not only the electrostatic potential, but also contains a magnetic contribution proportional to the $z$-component of periodic part of the vector potential $\mathbf{A}_\text{p}$. Detailed simulations show that the differential phase contrast imaging contains information about projected microscopic magnetic fields. The average strength of the magnetic signal is typically well below 1\% when compared to the electric signal component. An approach for its extraction has been described.

\begin{acknowledgments}
We acknowledge Swedish Research Council for financial support. The simulations were performed on resources provided by the Swedish National Infrastructure for Computing (SNIC) at the NSC center (computer clusters Triolith and Tetralith). AL acknowledges funding from the German Research Foundation (SPP 2137, LU 2261/2-1).
\end{acknowledgments}

\appendix

\section{Proof that DPC pattern remains conservative even in presence of magnetic field}\label{sec:proof}

In the text below we assume that the thickness is constant, motivated by atomic size of electron beams and assuming a small lateral extent of studied region of the sample.

To prove the conservative nature we show that the curl of the DPC pattern vanishes. Taking the curl of the center of mass of the diffraction patterns for different STEM-probe positions reads
\begin{align}
\nabla_\mathbf{R} \times \mathbf{\bar{E}}^\otimes = &  \nabla_\mathbf{R} \times \int \mathbf{\bar{E}}(\mathbf{r}_\perp) I_0 (\mathbf{r}_\perp - \mathbf{R}) \dd^2 \mathbf{r}_\perp \nonumber \\ 
= & \int \nabla_\mathbf{R} \times  \left[ \mathbf{\bar{E}}(\mathbf{r}_\perp) I_0 (\mathbf{r}_\perp - \mathbf{R}) \right] \dd^2 \mathbf{r}_\perp \nonumber \\
= & - \int  \mathbf{\bar{E}}(\mathbf{r}_\perp) \times \nabla_\mathbf{R} I_0 (\mathbf{r}_\perp - \mathbf{R}) \dd^2 \mathbf{r}_\perp \nonumber \\
= & \int  \mathbf{\bar{E}}(\mathbf{r}_\perp) \times \nabla_{\mathbf{r}_\perp} I_0 (\mathbf{r}_\perp - \mathbf{R}) \dd^2 \mathbf{r}_\perp \nonumber \\
= & - \int  \nabla_{\mathbf{r}_\perp} \times (\mathbf{\bar{E}}(\mathbf{r}_\perp) I_0 (\mathbf{r}_\perp - \mathbf{R})) \dd^2 \mathbf{r}_\perp \nonumber = 0, 
\end{align}
where it was used that $\nabla \times (\mathbf{a}(\mathbf{r}) f(\mathbf{r})) = -\mathbf{a}(\mathbf{r}) \times \nabla f(\mathbf{r})$ for a curl-free $\mathbf{a}(\mathbf{r})$ (holding for static electric fields). The last integral over $\dd^2 \mathbf{r}_\perp$ is zero for any sufficiently quickly decaying wave-function due to Stokes theorem.

Similarly one can deal with curl of the magnetic terms in Eq.~\ref{eq:pR}. 
We first observe that
\begin{equation}
  \nabla_\mathbf{R} \times ed(-\bar{B}_y^\otimes,\bar{B}_x^\otimes) = ed \nabla_\mathbf{R} \cdot \mathbf{\bar{B}}^\otimes
\end{equation}
and then analogically we show that
\begin{equation}
  \nabla_\mathbf{R} \cdot \mathbf{\hat{B}}^\otimes = \int \mathbf{\bar{B}}(\mathbf{r}_\perp) \cdot \nabla_{\mathbf{r}_\perp} I_0(\mathbf{r}_\perp - \mathbf{R}) \dd^2 \mathbf{r}_\perp = 0
\end{equation}
because
\begin{eqnarray}
\lefteqn{\nabla_{\mathbf{r}_\perp} \cdot \left( \mathbf{\bar{B}}(\mathbf{r}) I_0 (\mathbf{r} - \mathbf{R}) \right) = }\nonumber \\
& = & I_0 (\mathbf{r} - \mathbf{R}) \nabla_{\mathbf{r}_\perp} \cdot \mathbf{\bar{B}}(\mathbf{r}) \nonumber + \left(  \nabla_{\mathbf{r}_\perp} I_0 (\mathbf{r} - \mathbf{R}) \right) \cdot \mathbf{\bar{B}}(\mathbf{r})
\end{eqnarray}
where integral of the left-hand side over $\dd \mathbf{r}_\perp$ is zero due to two-dimensional divergence theorem and the quick decay of $I_0$ for large $|\mathbf{r}_\perp|$. Maxwell equation $\nabla \cdot \mathbf{B}=0$ in combination with periodicity of $\mathbf{B}$ implies that the integral containing $\nabla_{\mathbf{r}_\perp} \cdot \mathbf{\bar{B}}(\mathbf{r})$ is zero as well. This concludes the proof that the physical atomic resolution STEM-DPC pattern remains curl-free (i.e., conservative) even in the presence of magnetic fields.

\end{document}